%% file: LiftedCLMRIdentification.tex
\documentclass[5p,times,twocolumn]{elsarticle}
\usepackage{setspace}
\usepackage{pgfornament}
\usepackage{wrapfig}
\usepackage{mathrsfs}  
\usepackage{subcaption}
\usepackage{myStyle}
\usepackage{eso-pic}
\newif\ifthesismode
\thesismodefalse 
\renewcommand{\citet}{\cite}
\renewcommand{\citep}{\cite}

\newcommand{\smallMin}{\raisebox{-0.5pt}{\ensuremath{\scalebox{0.65}[0.65]{$-$}}}}
\AddToShipoutPictureBG*{%
	\AtPageUpperLeft{%
		\setlength\unitlength{1in}%
		\hspace*{\dimexpr0.5\paperwidth\relax}
		\makebox(0,-0.75)[c]{\begin{tabular}{c c}
				Max van Haren et al,	\textit{Lifted Frequency-Domain Identification of Closed-Loop Multirate Systems:} \\ \textit{Applied to Dual-Stage Actuator Hard Disk Drives,} Mechatronics, 108:103311, uploaded \today  \\
		\end{tabular}}
}}
\begin{document}
	\begin{frontmatter}
	\title{Lifted Frequency-Domain Identification of Closed-Loop Multirate Systems: \\ Applied to Dual-Stage Actuator Hard Disk Drives\tnoteref{label1}}
	\tnotetext[label1]{This research has received funding from the ECSEL Joint Undertaking under grant agreement 101007311 (IMOCO4.E), which receives support from the European Union Horizon 2020 research and innovation programme.}
\author[1]{Max van Haren\corref{cor1}}
\cortext[cor1]{Corresponding author.}
\ead{m.j.v.haren@tue.nl}
\author[2]{Masahiro Mae}
\author[3,1]{Lennart Blanken}
\author[1,4]{Tom Oomen}
\affiliation[1]{organization={Department of Mechanical Engineering, Control Systems Technology Section, Eindhoven University of Technology},
	addressline={Groene Loper 5},
	city={Eindhoven},
	postcode={5612 AE}, 
	country={The Netherlands}.}
\affiliation[2]{organization={Department of Electrical Engineering and Information Systems, The University of Tokyo},
	city={Tokyo},
	postcode={113-8656}, 
	country={Japan}.}
\affiliation[3]{organization={Sioux Technologies},
	addressline={Esp 130},
	city={Eindhoven},
	postcode={5633 AA}, 
	country={The Netherlands}.}
\affiliation[4]{organization={Delft Center for Systems and Control, Delft University of Technology},
	addressline={Mekelweg 2},
	city={Delft},
	postcode={2628 CN}, 
	country={The Netherlands}.}
	\begin{abstract}		
\input{abstract}
\end{abstract} 
\end{frontmatter}
\input{Introduction}
\input{ProblemDefinition}
\input{Approach}
\input{Simulation}
\input{conclusions}
\bibliographystyle{elsarticle-num}
\bibliography{../../library}

\end{document}

%% file: abstract.tex
Frequency-domain representations are crucial for the design and performance evaluation of controllers in multirate systems, specifically to address intersample performance. The aim of this \manuscript is to develop an effective frequency-domain system identification technique for closed-loop multirate systems using solely slow-rate output measurements. By indirect identification of multivariable time-invariant representations through lifting, in combination with local modeling techniques, the multirate system is effectively identified.	The developed method is capable of accurate identification of closed-loop multirate systems within a single identification experiment, using fast-rate excitation and inputs, and slow-rate outputs. Finally, the developed framework is validated using a benchmark problem consisting of a multivariable dual-stage actuator from a hard disk drive, demonstrating its applicability and accuracy.

%% file: Introduction.tex
\section{Introduction}
Multirate sampling is becoming more common in mechatronics as increasing complexity results in multiple systems, sensors, and actuators with different sampling rates being interconnected. Common examples of multirate systems include sampled-data control systems \citep{Chen1995} and hard disk drives \citep{Bashash2019,Atsumi2023a}. Traditional system identification techniques for multirate closed-loop systems are generally hampered due to the lack of Linear Time Invariant (LTI) properties of the closed-loop. 

Frequency Response Functions (FRFs) are widely used for representing mechatronic systems, and are fast, accurate, and cost-effective \citep{Pintelon2012}. They are suitable for closed-loop systems \cite{Evers2020} and directly enable the analysis of stability, performance, and robustness \citep{Skogestad2005}. Traditionally, FRFs were measured using random broadband excitation signals in combination with averaging and windowing techniques, having widespread applications in mechatronics such as \citet{VandeWal2002}. The advantages of periodic signals, including multisines, were recognized following the use of random excitation signals, as demonstrated for mechatronics in \citet{Oomen2014}. More recently, FRF identification has been improved through the use of local modeling techniques \citep[Chapter~7]{Pintelon2012}, particularly in suppressing the transient behavior. Furthermore, local modeling directly identifies multivariable systems using significantly less data, as seen in mechatronic applications such as vibration isolation systems \citep{Voorhoeve2018} and deformable mirrors \citep{Tacx2024}.

The development of identification methods for FRFs of multirate systems has been limited. Several approaches are developed to identify an FRF beyond the Nyquist frequency of a slow-rate output for open-loop single-input single-output systems using swept sines \citep{Ehrlich1989} or local modeling techniques \citep{VanHaren2023a}. While parametric identification techniques for multirate systems such as \citet{Ding2011} have been alternatively developed, the non-parametric FRFs frequently used in mechatronics are not directly identified. These methods do not account for closed-loop and multivariable effects, or focus on parametric models, and therefore cannot identify multivariable FRFs for closed-loop multirate systems.

Although FRF identification has significantly progressed, an effective multivariable FRF identification technique for closed-loop multirate systems beyond the Nyquist frequency of a slow-rate output is still lacking. In this \manuscript, FRFs of multirate systems are effectively identified through synergistic use of time-invariant representations by lifting and local modeling techniques. The developed method is directly capable of identifying multi-input multi-output systems beyond the Nyquist frequency of a slow-rate output, and is highly suitable for lightly-damped systems. The contributions of this \manuscript include the following.
\begin{enumerate}
\item[C1)] Indirect FRF identification of closed-loop multirate systems through lifting to a multivariable time-invariant representation, enabling systematic identification beyond the Nyquist frequency of a slow-rate output.
\item[C2)] Efficient single-experiment FRF identification of the time-invariant representations through local modeling.
\item[C3)] Validation of the developed framework on an benchmark challenge of a dual-stage actuator of a hard disk drive. 
\end{enumerate}
\paragraph*{Notation}
Fast-rate signals are denoted by subscript $h$, and slow-rate signals with subscript $l$, which have been downsampled by a factor $\fac\in\mathbb{Z}_{>0}$, with integers $\mathbb{Z}$. Fast-rate and slow-rate signals consists of respectively $N$ and $M=\frac{N}{\fac}$ data points. The $N$-points and $M$-points Discrete Fourier Transform (DFT) for finite-time fast-rate and slow-rate signals is given by
\begin{equation}
	\label{CLMR:eq:DFT4}
		\begin{aligned}
			X_h(k) &= \sum_{\dt=0}^{N-1} x_h(\dt) e^{\smallMin j\omega_k \dt \tsh}, \\
			X_l(k) &= \sum_{\ldt=0}^{M-1} x_l(\ldt) e^{\smallMin j \omega_k \ldt \tsl } = \sum_{\dt=0}^{M-1} x_h(\dt\fac) e^{\smallMin j \omega_k \dt \tsl},
		\end{aligned}
\end{equation}
with complex variable $j=\sqrt{-1}$, sampling times $\tsh$ and $\tsl=\fac\tsh$, discrete-time indices for fast-rate signals $\dt\in \mathbb{Z}_{[0,N-1]}$ and slow-rate signals $\ldt\in \mathbb{Z}_{[0,M-1]}$, and frequency bin $k\in\mathbb{Z}_{[0,N-1]}$, that relates to the frequency grid
\begin{equation}
	\label{CLMR:eq:omegak}
	\omega_k=\frac{2\pi k}{N \tsh}=\frac{2\pi k}{M \tsl} \in \mathbb{R}_{\left[0,{2\pi}/{\tsh}\right)}.
\end{equation}

%% file: ProblemDefinition.tex
\section{Problem Formulation}
In this section, a motivating application and the identification setting are shown for multirate closed-loop identification. Additionally, a frequency-domain analysis of the multirate closed-loop is shown. Finally, the problem treated in this \manuscript is defined.

\subsection{Motivating Application}
\label{CLMR:sec:motivatingApplication}
\begin{figure}[tb]
	\centering
	\ifthesismode
	\includegraphics{HDDTotalThesis.pdf}
	\else
	\includegraphics{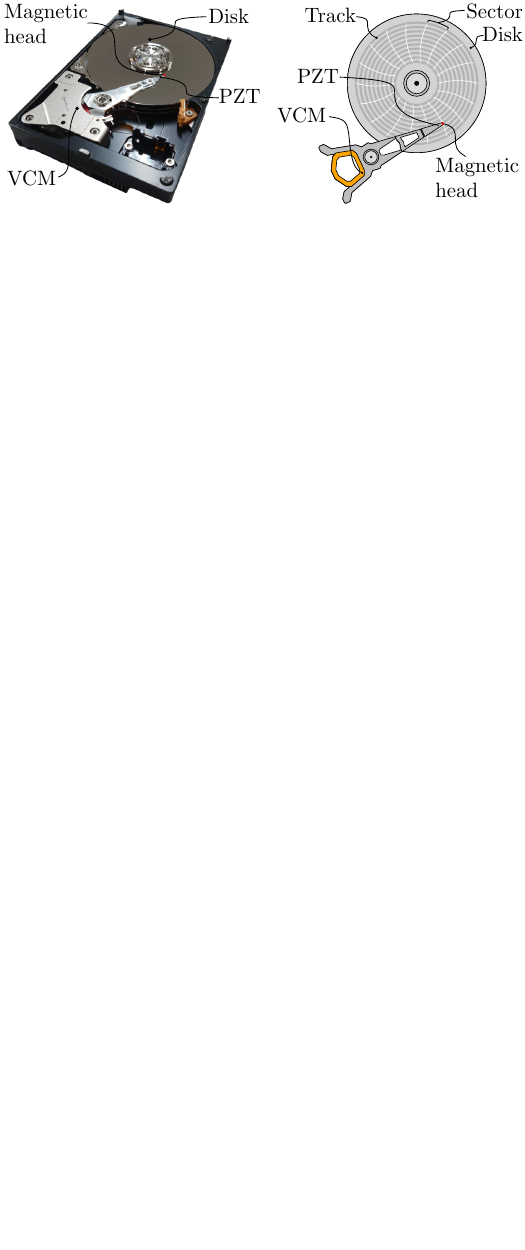}
	\fi
	\caption{(Left:) Photograph of hard disk drive with dual-stage actuator, consisting of a VCM and a PZT actuator. (Right:) Schematic overview of hard disk drive with dual-stage actuator.}
	\label{CLMR:fig:HDD}
\end{figure}

The problem addressed in this \manuscript is directly motivated by a benchmark dual-stage actuator hard disk drive challenge \citep{Atsumi2023a}, which has seen various applications in control, including \citet{SuryaPrakash2023, Wang2023, Mae2023a}. The dual-stage actuator hard disk drive shown in \figRef{CLMR:fig:HDD} exemplifies a closed-loop multirate system with slow-rate position measurements.
The dual-stage actuator system consists of a Voice Coil Motor (VCM) and a lead-zirconate-titanate (PbZrTi) piezoelectric actuator, denoted with PZT. The goal of the dual-stage actuator system is to minimize the tracking error of the magnetic head with respect to a track on the hard disk, as shown in \figRef{CLMR:fig:HDD}. The output is sampled at a slow-rate since the magnetic head position is determined based on a limited amount of sectors written on the disk. The PZT actuator has high stiffness, and therefore high-frequent resonance modes. By sampling the input to the actuators at a higher rate compared to the output, it enables active control of the resonances beyond the Nyquist frequency of the slow-rate output by means of feedback control \citep{Atsumi2023a}. The dual-stage actuator, with higher input sampling rate than output sampling rate, directly motivates the development of identification techniques for fast-rate systems in multirate closed-loop using slow-rate outputs.
\subsection{Identification Setting}
\begin{figure}[tb]
	\centering
	\includegraphics{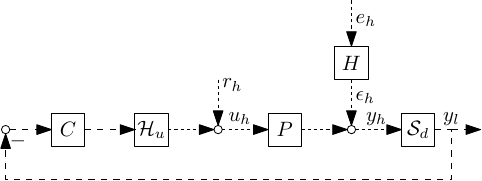}
	\caption{Multirate feedback loop considered, where fast-rate system $P$ is to be identified.}
	\label{CLMR:fig:IdentificationSetting}
\end{figure}

The goal is to identify a non-parametric FRF of fast-rate system $P$ having $n_y$ outputs and $n_u$ inputs, using slow-rate output $y_l(\ldt)\in\mathbb{R}^{n_y}$, fast-rate excitation signal $r_h(\dt)\in\mathbb{R}^{n_u}$, and fast-rate input $u_h(\dt)\in\mathbb{R}^{n_u}$, as shown in \figRef{CLMR:fig:IdentificationSetting}.
The set of systems $P$ that is considered is described by the LTI discrete-time rational transfer function
\begin{equation}
	\label{CLMR:eq:RationalSystem}
	\begin{aligned}
		P\left(q\right) = \frac{B\left(q\right)}{A\left(q\right)} 
	\end{aligned}
\end{equation}
where $q^{-1}$ denotes the lag operator $q^{-1}x(\dt) = x(\dt-1)$, and $A(q)$ and $B(q)$ are polynomial matrices in $q$. The interpolator $\mathcal{H}_u$ consists of a zero-order hold filter and an upsampler $\mathcal{H}_u=\mathcal{I}_{ZOH}(q) \mathcal{S}_u$, with upsampler \citep{Vaidyanathan1993}
\begin{equation}
	\begin{aligned}
		\nu_h(\dt) = \mathcal{S}_u\nu_l(\ldt) = \begin{cases}
			\nu_l\left(\frac{\dt}{\fac}\right) & \text{for } \frac{\dt}{\fac}\in\mathbb{Z},\\
			0& \text{for } \frac{\dt}{\fac}\notin\mathbb{Z}.
		\end{cases}
	\end{aligned}
\end{equation}
The zero-order hold filter is defined by
\begin{equation}
	\label{CLMR:eq:ZOH}
	\begin{aligned}
		\mathcal{I}_{ZOH}(q) = \sum_{f=0}^{\fac-1}q^{-f}.
	\end{aligned}
\end{equation}
The downsampler $\mathcal{S}_d$ is described by \citep{Vaidyanathan1993}
\begin{equation}
	\label{CLMR:eq:downsampler}
	\begin{aligned}
			\nu_l(\ldt) = \mathcal{S}_d\nu_h(\dt) = \nu_h(\fac \ldt).
	\end{aligned}
\end{equation}
\subsection{Frequency-Domain Analysis of Multirate Closed-Loop}
The fast-rate output $y_h$ of the closed-loop system in \figRef{CLMR:fig:IdentificationSetting} for excitation signal $r_h$ and noise $\epsilon_h=He_h$, with $e_h$ zero-mean, independent, and identically distributed noise, is given by
\begin{equation}
	\label{CLMR:eq:ClosedLoopIO}
	\begin{aligned}
		y_h = \left(I+P\mathcal{H}_uC\mathcal{S}_d\right)^{-1}\left(Pr_h+\epsilon_h\right).
	\end{aligned}
\end{equation}
Taking the DFT \eqref{CLMR:eq:DFT4} on both sides of \eqref{CLMR:eq:ClosedLoopIO}, the output is described in the frequency-domain by \citep{Oomen2007}
\begin{equation}
	\label{CLMR:eq:ClosedLoopIO_DFT}
	\ifthesismode
		\begin{aligned}
			&Y_h(k) \!=\! P(\freq_k)R_h(k)\!+\!E_h(k)\smallMin P(\freq_k)\mathcal{I}_{ZOH}(\freq_k) \left(
			I\!+\!C(\freq_k)P_l(\freq_k)
			\right)^{-1}\\
			\cdot &C(\freq_k)\left(\frac{1}{\fac}\sum_{f=0}^{\fac-1}P(\freq_{k+fM})R_h(k+fM)+E_h(k+fM)\right)+T_Y(\freq_k),
		\end{aligned}
	\else
	\resizebox{\linewidth}{!}{$ \displaystyle
	\begin{aligned}
		&Y_h(k) \!=\! P(\freq_k)R_h(k)\!+\!E_h(k)\smallMin P(\freq_k)\mathcal{I}_{ZOH}(\freq_k) \left(
		I\!+\!C(\freq_k)P_l(\freq_k)
		\right)^{-1}\\
		\cdot &C(\freq_k)\left(\frac{1}{\fac}\sum_{f=0}^{\fac-1}P(\freq_{k+fM})R_h(k+fM)+E_h(k+fM)\right)+T_Y(\freq_k),
	\end{aligned}$}
\fi
\end{equation}
with generalized frequency variable $\freq_k = e^{-j\omega_k \tsh}$, transient contribution $T_Y\left( \freq_k\right) \in \mathbb{C}^{n_y}$, which includes leakage and occurs due to finite-length signals \citep{Schoukens2009}, and 
\begin{equation}
	\begin{aligned}
		P_l(\freq_k) = \frac{1}{\fac} \sum_{f=0}^{\fac-1} P(\freq_{k+fM})\mathcal{I}_{ZOH}(\freq_{k+fM}).
	\end{aligned}
\end{equation}
The transient of the noise system $H$ is typically neglected, since it is negligible compared to its steady-state response \citep[Section~6.7.3.4]{Pintelon2012}. The fast-rate output $y_h$ in \eqref{CLMR:eq:ClosedLoopIO} and \eqref{CLMR:eq:ClosedLoopIO_DFT} is downsampled with \eqref{CLMR:eq:downsampler}, as shown in \figRef{CLMR:fig:IdentificationSetting}, having DFT \citep{Vaidyanathan1993}
\begin{equation}
	\label{CLMR:eq:FDomainSlowSampledOutput1}
	\begin{aligned}
		Y_l(k)=\frac{1}{\fac} \sum_{f=0}^{\fac-1} Y_h(k+fM).
	\end{aligned}
\end{equation}
\subsection{Problem Definition}
The slow-rate output \eqref{CLMR:eq:FDomainSlowSampledOutput1} is influenced by multiple contributions of the excitation signal $r_h$ due to the multirate closed-loop and the downsampling operation, as shown in \eqref{CLMR:eq:ClosedLoopIO_DFT} and \eqref{CLMR:eq:FDomainSlowSampledOutput1}. Consequently, LTI identification techniques that assume the frequency separation principle are not capable of identifying the multirate system.

Therefore, the problem considered in this \manuscript is as follows. Given fast-rate excitation signal $r_h$, fast-rate input to the system $u_h$, and slow-rate output of the system $y_l$ as shown in \figRef{CLMR:fig:IdentificationSetting}, identify the fast-rate FRF $P(\freq_k)$ for all $k\in\mathbb{Z}_{[0,N-1]}$.

%% file: Approach.tex
\section{Lifted Closed-Loop Identification}
\label{CLMR:sec:method}
In this section, multirate closed-loop systems are identified through lifting them to multivariable time-invariant representations. Furthermore, the time-invariant representations are indirectly identified in a single identification experiment. The developed approach is then summarized in a procedure.
\subsection{Multirate Identification through Lifting}
The fast-rate system $P$ is identified using slow-rate output measurement through lifting the input-output behavior. Given a signal $x_h(\dt)\in\mathbb{R}^{n_x}$, the lifted signal $\underline{x}(\ldt)\in \mathbb{R}^{\fac n_x}$ is defined as
\begin{equation}
	\label{CLMR:eq:liftsig}
	\begin{aligned}
		\underline{x}(\ldt) &= \mathcal{L} x_h(\dt) \\
		&= \begin{bmatrix}	x_h^\top(\ldt\fac) & x_h^\top(\ldt\fac+1) & \cdots & x_h^\top(\ldt\fac+\fac-1).
		\end{bmatrix}^\top 
	\end{aligned}
\end{equation}
For LTI system $P$, the FRF of lifted system $\underline{P}=\mathcal{L}P\mathcal{L}^{-1}$ is given by \citep{Bittanti2000}
\begin{equation}
	\label{CLMR:eq:LiftedP}
	\begin{aligned}
		\resizebox{\linewidth}{!}{$ \displaystyle
		\underline{P}\left(\freq_k^\fac\right) = \begin{bmatrix}
			P^{(0)}\left(\freq_k^\fac\right) & \freq_k^\fac P^{(\fac\smallMin1)}\left(\freq_k^\fac\right) & \freq_k^\fac P^{(\fac\smallMin2)}\left(\freq_k^\fac\right) & \!\!\cdots\!\! & \freq_k^\fac P^{(1)}\left(\freq_k^\fac\right) \\
			P^{(1)}\left(\freq_k^\fac\right) & P^{(0)}\left(\freq_k^\fac\right) & \freq_k^\fac P^{(\fac\smallMin1)}\left(\freq_k^\fac\right) & \!\!\cdots\!\! & \freq_k^\fac P^{(2)}\left(\freq_k^\fac\right) \\
			\vdots & \vdots & \vdots & \!\!\ddots\!\! & \vdots \\
			P^{(\fac\smallMin1)}\left(\freq_k^\fac\right) & P^{(\fac\smallMin2)}\left(\freq_k^\fac\right) & P^{(\fac\smallMin3)}\left(\freq_k^\fac\right) & \!\!\cdots\!\! & P^{(0)}\left(\freq_k^\fac\right)
		\end{bmatrix},$}
	\end{aligned}
\end{equation}
where $\underline{P}\left(\freq_k^\fac\right) \in \mathbb{C}^{\fac n_y \times \fac n_u}$ is described using
\begin{equation}
	\label{CLMR:eq:LiftedPEntries}
	\begin{aligned}
		P^{(i)}\left(\freq_k^\fac\right)=\frac{\freq_k^i}{\fac} \sum_{f=0}^{\fac-1} P\left(\phi^f\freq_k \right) \phi^{f i}, \quad \phi=e^{\frac{2\pi j}{\fac}}.
	\end{aligned}
\end{equation}
Clearly, from the distinctive structure of $\underline{P}$ in \eqref{CLMR:eq:LiftedP} and its entries described with \eqref{CLMR:eq:LiftedPEntries}, the original system $P$ is recovered using the first row of $\underline{P}$ as
\begin{equation}
	\label{CLMR:eq:RecoverP}
	\begin{aligned}
		P\left( \freq_{k}\right) =\underline{P}_{[1,1]}\left(\freq_k^\fac\right)+ e^{j\omega_k \tsl} \sum_{f=1}^{F-1}\left(e^{-j\omega_k \tsh}\right)^{-\left(\fac-f\right)}\underline{P}_{[1,f+1]}\left(\freq_k^\fac\right),
	\end{aligned}
\end{equation}
where $\underline{P}_{[i,i]}\left(\freq_k\right) \in\mathbb{C}^{n_y \times n_u}$ indicates the $[i,i]^{\textrm{th}}$ block of $\underline{P}$. 

The first row of the lifted system $\underline{P}$ is described by using the slow-rate output measurements $y_l$ through $\mathcal{S}_d\mathcal{L}^{-1} = \begin{bmatrix}	I & 0 & \cdots & 0\end{bmatrix}$ \citep[Section~8.3]{Chen1995}, resulting in the LTI behavior
\begin{equation}
	\label{CLMR:eq:LTI_OL}
	\begin{aligned}
		y_l &= \mathcal{S}_d \left(P u_h+\epsilon_h\right) = \mathcal{S}_d\mathcal{L}^{-1} \left(\mathcal{L}P\mathcal{L}^{-1} \underline{u}+\mathcal{L}\mathcal{L}^{-1} \underline{\epsilon}\right) \\
		&= \begin{bmatrix}	I & 0 & \cdots & 0\end{bmatrix} \left(\underline{P} \; \underline{u}+\underline{\epsilon}\right).
	\end{aligned}
\end{equation}
Therefore, by multivariable LTI identification of $\begin{bmatrix}	I & 0 & \cdots & 0\end{bmatrix} \underline{P}$ using slow-rate outputs $y_l$ shown in \eqref{CLMR:eq:LTI_OL}, the fast-rate system $P$ is recovered by utilizing \eqref{CLMR:eq:RecoverP}. However, the input $\underline{u}$ and the noise $\underline{\epsilon}$ are correlated, since the system is operating in multirate closed-loop, resulting in a biased FRF estimate \citep[Section~2.6]{Pintelon2012}.
\subsection{Indirect Closed-Loop Identification}
By performing indirect identification, the bias observed for direct closed-loop identification of the first row of $\underline{P}$ is effectively avoided \citep[Section~2.6.4]{Pintelon2012}. First, lift the excitation signal $\underline{r}=\mathcal{L}r_h$ and describe the lifted input $\underline{u}$ in closed-loop as
\begin{equation}
	\label{CLMR:eq:LiftedInputSystem}
	\begin{aligned}
\mathcal{L}u_h = \underline{u} = &\underline{S}\: \underline{r} + \underline{CS}\: \underline{\epsilon} \\
		= &\mathcal{L} \left(I+\mathcal{H}_u C \mathcal{S}_d P\right)^{-1}\mathcal{L}^{-1} \underline{r} \\
		+&\mathcal{L} \left(I+\mathcal{H}_u C \mathcal{S}_d P\right)^{-1}\mathcal{H}_uC \mathcal{S}_d \mathcal{L}^{-1}\underline{\epsilon},
	\end{aligned}
\end{equation}
with lifted sensitivity $\underline{S}$ and lifted control sensitivity $\underline{CS}$. Similarly, the slow-rate output $y_l$ in closed-loop is given by
\begin{equation}
	\label{CLMR:eq:LiftedSlowOutput}
	\begin{aligned}
	y_l &= \mathcal{S}_d  \left(I+P\mathcal{H}_u C \mathcal{S}_d \right)^{-1}\left(P\mathcal{L}^{-1} \underline{r}+\mathcal{L}^{-1} \underline{\epsilon}\right) \\
	&= \mathcal{S}_d \mathcal{L}^{-1}\mathcal{L} \left(I+P\mathcal{H}_u C \mathcal{S}_d \right)^{-1}\left(P\mathcal{L}^{-1} \underline{r}+\mathcal{L}^{-1} \underline{\epsilon}\right) \\
	& = \begin{bmatrix}
		I & 0 & \cdots & 0
	\end{bmatrix} \left(\underline{PS}\:\underline{r}+\underline{S}\: \underline{\epsilon} \right) \\
:&=\underline{PS}_{[1,:]} \underline{r}+\underline{S}_{[1,:]} \underline{\epsilon}
\end{aligned}
\end{equation}
with lifted process-sensitivity $\underline{PS}$. The first row of $\underline{P}$ is determined indirectly by
\begin{equation}
	\label{CLMR:eq:FirstRowPbar}
	\begin{aligned}
		\begin{bmatrix}
		I & 0 & \cdots & 0
	\end{bmatrix} \underline{P} &= \begin{bmatrix}
I & 0 & \cdots & 0
\end{bmatrix} \underline{PS}\: \underline{S}^{-1} =\underline{PS}_{[1,:]} \underline{S}^{-1}.
	\end{aligned}
\end{equation}
Subsequently, the system $P$ is determined with \eqref{CLMR:eq:RecoverP}. 

In summary, the original system $P$ is recovered via the first row of its lifted representation $\underline{P}$, which is determined using the lifted sensitivity $\underline{S}$ and the first row of the lifted process sensitivity $\underline{PS}_{[1,:]}$. The lifted (process) sensitivity can be indirectly identified using multivariable identification techniques utilizing the slow-rate output $y_l$ and the fast-rate signals $r_h$ and $u_h$.
\subsection{Multivariable Identification through Local Modeling}
\label{CLMR:sec:LocalModeling}
In this section, the lifted closed-loop systems are effectively identified using multivariable local rational modeling \citep{Voorhoeve2018}. In a local frequency window $r\in\mathbb{Z}_{[-\wsize,\wsize]}$, the DFTs of the lifted input to the system $\underline{u}$ and the slow-rate output $y_l$ from \eqref{CLMR:eq:LiftedInputSystem} and \eqref{CLMR:eq:LiftedSlowOutput} are approximated by neglecting the noise as
\begin{equation}
	\label{CLMR:eq:EstOutput}
	\begin{aligned}
		\widehat{Z}(k+r) \! := \!\begin{bmatrix}
			\widehat{\underline{U}}(k+r) \\
			\widehat{Y}_l(k+r)
		\end{bmatrix} &\!=\! \begin{bmatrix}
		\widehat{\underline{S}}\left( \freq_{k+r}\right) \\
		\widehat{\underline{PS}}_{[1,:]}\left( \freq_{k+r}\right)
	\end{bmatrix}\underline{R}(k+r)\! + \begin{bmatrix}
	\widehat{T}_{\underline{U}}\left(\freq_{k+r}\right) \\
	\widehat{T}_{Y_l}\left(\freq_{k+r}\right)
\end{bmatrix}\\
:&= \widehat{G}\left( \freq_{k+r}\right)\underline{R}(k+r)+ \widehat{T}_G\left( \freq_{k+r}\right),
	\end{aligned}
\end{equation}
with multivariable system $\widehat{G}\left( \freq_{k+r}\right)\in\mathbb{C}^{n_u\fac+n_y\times n_u\fac}$ and transient $\widehat{T}_G\left( \freq_{k+r}\right) \in\mathbb{C}^{n_u\fac+n_y}$. The multivariable system and transient are modeled using the local rational models
\begin{equation}
	\label{CLMR:eq:localModel}
	\begin{aligned}
		\widehat{G}\left( \freq_{k+r}\right)  &= D\left( \freq_{k+r}\right)^{-1}N\left( \freq_{k+r}\right), \\ 
		\widehat{T}_G\left( \freq_{k+r}\right)  &= D\left( \freq_{k+r}\right)^{-1}M\left( \freq_{k+r}\right)
	\end{aligned}
\end{equation}
with 
\begin{equation}
	\label{CLMR:eq:LocalDenNum}
	\begin{aligned}
		N\left( \freq_{k+r}\right) &= \widehat{G}\left( \freq_{k+r}\right) +\sum_{s=1}^{R_n}N_s(k)r^s, \\
		M\left( \freq_{k+r}\right) &= \widehat{T}_G\left( \freq_{k+r}\right) +\sum_{s=1}^{R_m}M_s(k)r^s, \\
		D\left( \freq_{k+r}\right) &= I +\sum_{s=1}^{R_d}D_s(k)r^s. \\
	\end{aligned}
\end{equation}
The decision parameters $\widehat{G}\left( \freq_{k+r}\right)$, $\widehat{T}_G\left( \freq_{k+r}\right)$, $N_s(k)\in\mathbb{C}^{n_u\fac+n_y\times n_u\fac}$, $M_s(k)\in\mathbb{C}^{n_u\fac+n_y}$, and $D_s(k)\in\mathbb{C}^{n_u\fac+n_y\times n_u\fac+n_y}$ are determined by minimizing the weighted difference between approximated outputs \eqref{CLMR:eq:EstOutput} and measured outputs $Z(k+r)$, resulting in the linear least squares problem
\begin{equation}
	\label{CLMR:eq:CostFunction}
	\ifthesismode
		\begin{aligned}
			&\widehat{\Theta}(k) = \arg\min_\Theta \sum_{r=\smallMin \wsize}^{\wsize} \left\|\vphantom{\sum} D\left( \freq_{k+r}\right) \left(Z(k+r)-\widehat{Z}(k+r)\right)\right\|_2^2 \\
			&= \arg\min_\Theta \sum_{r=- \wsize}^{\wsize} \left\|\vphantom{\sum}
			D\left( \freq_{k+r}\right)Z(k+r)- N\left( \freq_{k+r}\right)\underline{R}(k+r)- M\left( \freq_{k+r}\right)
			\right\|_2^2,
		\end{aligned}
	\else
	\resizebox{\linewidth}{!}{$\displaystyle
	\begin{aligned}
		&\widehat{\Theta}(k) = \arg\min_\Theta \sum_{r=\smallMin \wsize}^{\wsize} \left\|\vphantom{\sum} D\left( \freq_{k+r}\right) \left(Z(k+r)-\widehat{Z}(k+r)\right)\right\|_2^2 \\
		 &= \arg\min_\Theta \sum_{r=- \wsize}^{\wsize} \left\|\vphantom{\sum}
		D\left( \freq_{k+r}\right)Z(k+r)- N\left( \freq_{k+r}\right)\underline{R}(k+r)- M\left( \freq_{k+r}\right)
		\right\|_2^2,
	\end{aligned}$}
\fi
\end{equation}
which has a unique closed-form solution \citep{Voorhoeve2018}.
\begin{remark}
	Note that an unweighted version of \eqref{CLMR:eq:CostFunction} can be minimized as well, either directly with non-linear optimization or by using iterative reweighted solutions such as the Sanathanan-Koerner algorithm \citep{Sanathanan1963}. However, these optimization strategies generally do not guarantee convergence to a global minimizer, and the weighted least-squares criterion \eqref{CLMR:eq:CostFunction} is especially effectively for practical applications \citep{Voorhoeve2018,Verbeke2020}.
\end{remark}
\begin{remark}
	\label{CLMR:rem:ExcitationSignal}
	Generally the cost function \eqref{CLMR:eq:CostFunction} has a unique closed-form solution only if the excitation signal $\underline{R}(k+r)$ is sufficiently 'rough' in the window $r\in\mathbb{Z}_{[-\wsize,\wsize]}$ \citep{Schoukens2009}. For example, orthogonal random-phase multisines \citep{Dobrowiecki2006} fulfill this condition.
\end{remark}
\subsection{Procedure Lifted Closed-Loop Identification}
The main results in \secRef{CLMR:sec:method} are summarized in Procedure~\ref{CLMR:proc:1}.
\begin{figure}[H]
	\hrule \vspace{1.5mm}\begin{proced}[Lifted closed-loop FRF identification of multirate system] \hfill \vspace{1mm} \hrule \vspace{-1mm}
		\label{CLMR:proc:1}
		\begin{enumerate}[itemsep=1pt]
			\item Construct excitation signal $r_h$.
			\item Excite system in \figRef{CLMR:fig:IdentificationSetting} with $r_h$ and record slow-rate output $y_l$ and fast-rate input to the system $u_h$.
			\item Lift fast-rate signals $u_h$ and $r_h$ into $\underline{u}=\mathcal{L} u_h$ and n$\underline{r}=\mathcal{L}r_h$ using \eqref{CLMR:eq:liftsig}.
			\item Take DFT of $\underline{r}$, $\underline{u}$, and $y_l$ using \eqref{CLMR:eq:DFT4}.
			\item  For frequency bins $k\in\mathbb{Z}_{[0,N-1]}$ identify $\widehat{P}\left(\freq_{k}\right)$ using the developed approach as follows.
			\begin{enumerate}[itemsep=1pt]
			\item Determine $\widehat{G}\left(\freq_{k}\right)$ in \eqref{CLMR:eq:EstOutput}, \eqref{CLMR:eq:localModel} and \eqref{CLMR:eq:LocalDenNum} by minimizing the linear least squares problem in \eqref{CLMR:eq:CostFunction}.
			\item Compute the first row of $\widehat{\underline{P}}\left(\freq_{k}\right)$ with \eqref{CLMR:eq:FirstRowPbar}, using $\widehat{\underline{S}}\left(\freq_{k}\right)$ and $\widehat{\underline{PS}}_{[1,:]}\left(\freq_{k}\right)$ contained in $\widehat{G}\left( \freq_{k}\right)$.
			\item Recover original fast-rate system $\widehat{P}(\freq_k)$ using \eqref{CLMR:eq:RecoverP}.
			\end{enumerate}
		\end{enumerate}
		\vspace{-5pt} 	\hrule \vspace{-9pt}
	\end{proced}
\end{figure}\vspace{-20pt}

%% file: Simulation.tex
\section{Benchmark Simulation}
In this section, the developed framework is validated on a benchmark dual-stage actuator hard disk drive challenge, leading to contribution C3. The benchmark challenge is introduced, followed by the results. 
\subsection{Dual-Stage Actuator Hard Disk Drive Identification Setup}
\begin{figure}[tb]
	\centering
	\ifthesismode
	\includegraphics{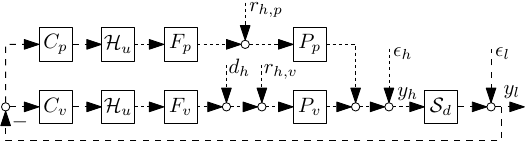}
	\else
	\includegraphics[width=\linewidth]{HDDLayout.pdf}
	\fi
	\caption{Feedback structure considered for identification of the PZT and VCM actuators $P_p$ and $P_v$ of the hard disk drive in \figRef{CLMR:fig:HDD}.}
	\label{CLMR:fig:HDDFB}
\end{figure}

\begin{figure}[tb]
	\centering
	\includegraphics{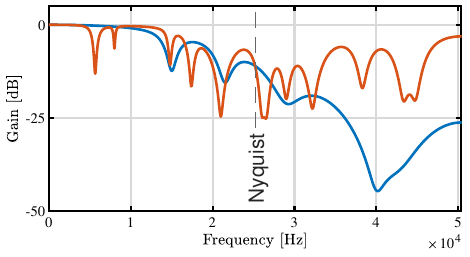}\vspace{-0.6em}
	\caption{The FRFs of $F_p$ \markerline{mblue} and $F_v$ \markerline{mred} show that the filters suppress dynamics beyond the Nyquist frequency of the slow-rate output $y_l$ \markerline{black}[dashed][.][0][0.5].}
	\label{CLMR:fig:FilterFRF}
\end{figure}

The benchmark challenge is a high-fidelity simulator of a dual-stage actuator hard disk drive developed by the IEEJ research committee of precision servo systems \citep{Atsumi2023a}, consisting of a VCM and PZT actuator as shown in \figRef{CLMR:fig:HDD} and \secRef{CLMR:sec:motivatingApplication}. The setup has $n_y=1$ outputs, which is the position of the magnetic head, and $n_u=2$ inputs for the VCM and PZT actuator. The system is in closed-loop control as shown in \figRef{CLMR:fig:HDDFB}.
The fast-rate noise $\epsilon_h$ is introduced by measurement noise and fan-induced vibration, while the VCM input is disturbed by rotational vibrations from other drives $d_h$. The slow-rate noise $\epsilon_l$, or repeatable run-out, is caused by the oscillations of target tracks on the disk. The disturbance spectra of $\epsilon_h$, $d_h$, and $\epsilon_{l}$ are seen in \citet{Atsumi2023a}. The system is under multirate closed-loop feedback control with controllers $C_p$ and $C_v$ sampled at \tsl, given by
\begin{equation}
	\begin{aligned}
		C_v(q) &= \frac{0.65 + 0.020 q^{-1} - 0.63 q^{-1}	}{1 - 1.4 q^{-1} + 0.51 q^{-2}}, \\
		C_p(q) &=  \frac{0.01346 + 0.01346 q^{-1}}{1 - 0.8825 q^{-1}}.	
	\end{aligned}
\end{equation}
In addition to the feedback controllers, the feedback loop contains two filters $F_p$ and $F_v$ sampled at \tsh, which consist of Notch filters to suppress high-frequent resonant behavior of the systems $P_p$ and $P_v$. The FRFs of $F_p$ and $F_v$ are seen in \figRef{CLMR:fig:FilterFRF}.
To design filters $F_p$ and $F_v$ such that they effectively suppress the resonant behavior, it is crucial to have an accurate model of fast-rate systems $P_p$ and $P_v$.

The excitation signals $r_p$ and $r_v$ for the PZT and VCM actuators are random phase multisines with a flat amplitude spectrum exciting all frequencies having root-mean-square values of $3.6\cdot10^{-9}$ and $8.0\cdot10^{-8}$. The maximum position of the PZT is 48.7 nm, and hence, does not exceed its stroke limit of 50 nm \citep{Atsumi2023a}. Additional settings during identification are seen in \tabRef{CLMR:tab:simValues}.
\begin{table}[tb]\vspace{-0.4em}
	\centering
	\caption{Identification settings.} \vspace{-0.4em}
	\label{CLMR:tab:simValues}
	\begin{tabular}{llll}
		\toprule
		\textbf{Variable}    & \textbf{Abbrevation} & \textbf{Value} & \textbf{Unit} \\
		\midrule
		Slow sampling time   & \tsl            &  \raisebox{1.5pt}{$\scriptstyle\frac{1}{50400}$}      & s   \\
		Fast sampling time   & \tsh            & \raisebox{1pt}{$\scriptstyle\frac{1}{100800}$}      & s   \\
		Downsampling factor& \fac & 2 & - \\
		Number of samples    & N               & 3600    & -    \\
		\bottomrule
	\end{tabular}
\end{table}
The polynomial degrees for local modeling in \eqref{CLMR:eq:LocalDenNum} are chosen as $R_n=R_m=R_d=3$, with window size $\wsize=30$. The denominator matrix $D_s(k)$ in \eqref{CLMR:eq:LocalDenNum} is parameterized using the multi-input single-output parameterization \citep{Voorhoeve2018} as
\begin{equation}
	\begin{aligned}
		D_s(k) = \operatorname{diag} \left(
		d_{s,1}(k),\; \cdots,\; d_{s,n_u\fac + n_y}(k)
		\right).
	\end{aligned}
\end{equation}
The total amount of decision parameters in \eqref{CLMR:eq:CostFunction} is given by $(n_u\fac+n_y) (n_u\fac(R_n+1)+R_m+1+R_d)=115$. Least-squares problem \eqref{CLMR:eq:CostFunction} has a unique closed-form solution since $\underline{U}(k+r)$ and $Y_l(k+r)$ contain $(\vphantom{n_u\fac+n_y}2\wsize+1)(n_u\fac+n_y)=305$ data points, and the excitation signal is a random-phase multisine, see \remRef{CLMR:rem:ExcitationSignal}.
\subsection{Identification Results}
\begin{figure}[tb]
	\centering
	\includegraphics{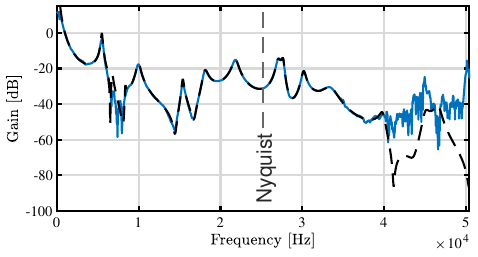}\vspace{-0.6em}
	\caption{The developed lifted closed-loop identification technique with local rational modeling \markerline{mblue} identifies the VCM $P_p(\freq_{k})$ accurately, even beyond the Nyquist frequency of the slow-rate output $y_l$ \markerline{black}[dashed][.][0][0.5].}
	\label{CLMR:fig:FRF_VCM}
	\includegraphics{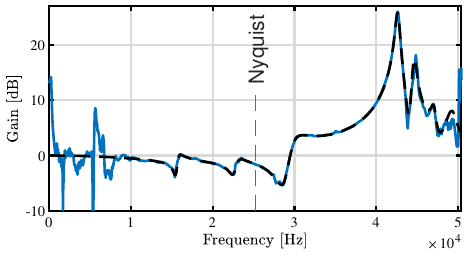}\vspace{-0.6em}
	\caption{The developed lifted closed-loop identification technique with local rational modeling \markerline{mblue} identifies the PZT $P_p(\freq_{k})$ accurately, even beyond the Nyquist frequency of the slow-rate output $y_l$ \markerline{black}[dashed][.][0][0.5].}
	\label{CLMR:fig:FRF_PZT}
\end{figure}

\begin{figure}[tb]
	\centering
	\includegraphics{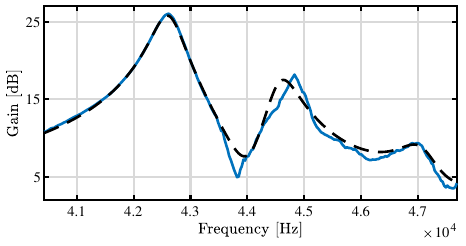}\vspace{-0.6em}
	\caption{The resonant dynamics of the PZT \markerline{black}[dashed] are accurately identified using the developed approach \markerline{mblue} beyond the Nyquist frequency of the slow-rate output $y_l$.}
	\label{CLMR:fig:FRF_PZT_Zoom}
\end{figure}

The identified FRFs of the VCM and PZT actuator $P_v$ and $P_p$ are seen in \figRef{CLMR:fig:FRF_VCM} and \figRef{CLMR:fig:FRF_PZT}.
Additionally, the resonant dynamics of the PZT around 44000 Hz are enlarged in \figRef{CLMR:fig:FRF_PZT_Zoom}.
The following observations are made with respect to the identified FRFs in Figures~\ref{CLMR:fig:FRF_VCM}, \ref{CLMR:fig:FRF_PZT}, and \ref{CLMR:fig:FRF_PZT_Zoom}.
\begin{itemize}[itemsep=1pt]
	\item While the system is operating in multirate closed-loop, the developed method identifies the FRFs of the VCM and PZT accurately, even beyond the Nyquist frequency of the slow-rate output $y_l$.
	\item The FRF quality is lower at low frequencies for the PZT. This occurs because the gain of the PZT at low frequencies is estimated simultaneously with its resonant behavior at high frequencies since these frequencies alias to the same frequency in $y_l$, resulting in a low "signal-to-signal" ratio. Similarly, the FRF quality of the VCM is lower at high frequencies.
\end{itemize} 
These observations show that the developed approach is capable of single-experiment identification of the multivariable fast-rate system while operating in multirate closed-loop using solely slow-rate position measurements of the actuator. 

%% file: conclusions.tex
\section{Conclusions}
The results in this \manuscript demonstrate the effective FRF identification of a dual-stage actuator hard disk drive operating in closed-loop beyond the Nyquist frequency of a slow-rate output. The key idea is to identify time-invariant representations of the multirate system through lifting the fast-rate signals. Additionally, the bias which is generally observed for direct identification of closed-loop systems is avoided through indirectly identifying the system. The time-invariant lifted representation of the multirate system is naturally multivariable, and therefore, local modeling techniques are utilized to effectively identify the multivariable system in a single identification experiment. Furthermore, the approach is directly suitable for multivariable system identification. The framework is validated using a benchmark dual-stage actuator hard disk drive, demonstrating accurate identification of the multi-input single-output actuator operating in multirate closed-loop utilizing only slow-rate position measurements. The developed approach is crucial in control design of multirate systems, including multivariable and closed-loop systems with slow-rate outputs.
\ifthesismode
\section*{Acknowledgments}
\else
\paragraph*{{Acknowledgments}}
\fi
The authors thank Leonid Mirkin and Roy S. Smith for their fruitful discussions that led to this \manuscript.
\ifthesismode
\else
\vspace{-0.6em}
\fi